\documentclass[amsmath, amsfonts, twocolumn, prb]{revtex4}
\usepackage{graphicx}
\usepackage{epsfig}
\usepackage{bm}
\usepackage{dcolumn}
\usepackage{amsmath}
\usepackage{amssymb}
\usepackage{color}

\newcommand{\Tr}{\mathop{\rm Tr}\nolimits}

\newcommand{\Th}{\mathop{\rm Th}\nolimits}

\newcommand{\arcth}{\mathop{\rm arctanh}\nolimits}

\begin{document}

\title{Josephson currents in chaotic quantum dots}

\author{Colin M. Whisler}
\affiliation{Department of Physics, University of Wisconsin-Madison, Madison, Wisconsin 53706, USA}

\author{Maxim G. Vavilov}
\affiliation{Department of Physics, University of Wisconsin-Madison, Madison, Wisconsin 53706, USA}

\author{Alex Levchenko}
\affiliation{Department of Physics, University of Wisconsin-Madison, Madison, Wisconsin 53706, USA}

\date{May 28, 2018}

\begin{abstract}
We study theoretically the Josephson current-phase relationship in a chaotic quantum dot coupled to superconductors by ballistic contacts. In this regime, strong proximity effect induces superconductivity in the quantum dot that leads to a significant modification in the electron density of states and formation of multiple sub-gaps. The magnitude of the resulting supercurrent depends on the phase difference of the superconducting order parameter in the leads and shows strongly anharmonic skewed behavior. We find that when the Thouless energy on the dot exceeds the superconducting energy gap, the second harmonic of the supercurrent becomes comparable in magnitude to the first harmonic. To address these effects on the technical level, we use the nonlinear $\sigma$-model Keldysh formalism in the framework of the circuit theory to compute dependence of the density of states, Josephson energy, and current on the superconducting phases in the leads. We analyze how these quantities change as a function of the Thouless energy and the superconducting gap. Finally, we briefly discuss sub-gap tail states, mesoscopic supercurrent fluctuations, weak localization correction, and also touch on anharmonicity of gatemon qubits with quantum dot Josephson junctions.  
\end{abstract}

\maketitle

\section{Introduction} 

The most profound fundamental properties and practical applications of superconductors are associated with  their behavior in the presence of spatial inhomogeneities on a mesoscopic scale. One example is given by the Josephson effect, which requires normal or insulating barrier between two superconducitng terminals~[\onlinecite{Kulik,Barone}]. Among different possible kinds of Josephson weak-links, the superconductor-normal metal-superconductor (SNS) junction is perhaps the most comprehensively studied system which reveals an incredibly rich physics~[\onlinecite{Birge,GKI}]. A particular model in this context is that of a chaotic cavity quantum dot (QD) where a piece of metallic grain is connected to the superconductors by means of point contacts that dominate the resistance of the structure in the normal state. It is well understood that the superconducting proximity effect in such a structure is governed by the processes of Andreev reflections of the normal electrons from the two NS boundaries.  An elegant way to describe this physics in technical terms is by means of random matrix and scattering matrix circuit theories~[\onlinecite{Beenakker,Nazarov}]. In this language one is able to relate the properties of the same structure in the normal and superconducting states, while circumventing the need for microscopic description of the structure in either of the two states. For that reason certain universal aspects of the proximity effect related to the single-particle density of states (DOS) and the Josephson current-phase-relationship (CPR) are known for such mesoscopic structures~[\onlinecite{Beenakker,Nazarov,Beenakker-3UJE,Belzig-Review}].      

\subsection{Overview}

In general, the properties of the S-QD-S junction are determined by the types of contacts, and also by the relationship between the superconducting energy gap $\Delta$ and the Thouless energy $E_{\Th}=G_T\delta/G_Q$. Here $\delta$ denotes the mean level spacing of the normal metal grain and $G_T\gg G_Q$ is the total conductance of the structure which is assumed to be large compared to the conductance quantum $G_Q=2e^2/h$. For simplicity we will discuss only the case of symmetric junctions both in terms of superconducting leads with identical energy gaps and properties of contacts. 

\begin{figure}
\includegraphics[width=\linewidth]{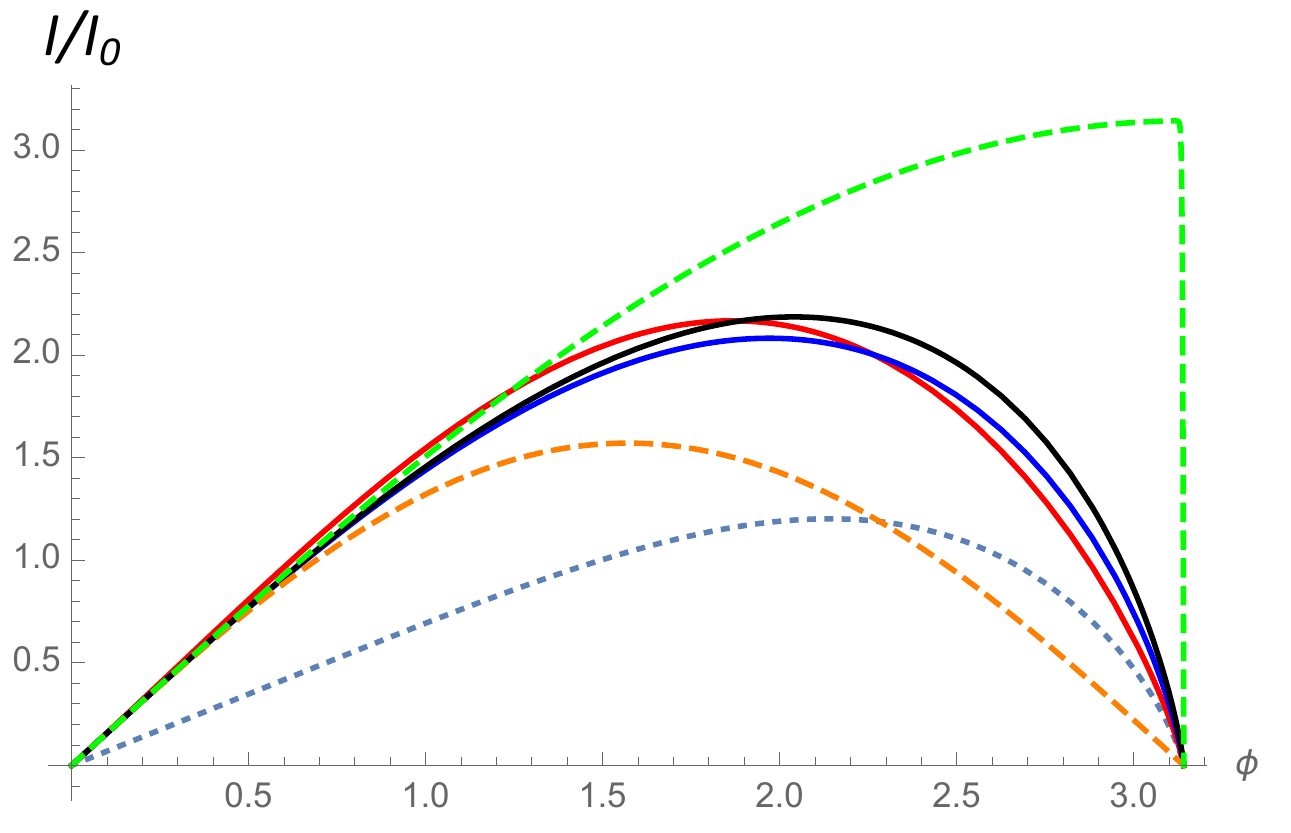} 
\caption{(Color online) Josephson current-phase relationships of various SNS junctions. Current is normalized in the units of $I_0=G_T\Delta/e$. The lowest in magnitude dotted line represents Eq. \eqref{I-SNS-1} plotted at $E_{\Th}=\Delta$. The middle three solid lines correspond to Eqs. \eqref{I-SINIS}--\eqref{I-SCCS}. The dashed lines serve as a reference to the conventional sinusoidal CPR of the tunnel junction as given by the Ambegaokar-Baratoff formula [\onlinecite{AB}]: $I(\phi)=(\pi G_T\Delta/2e)\sin(\phi)$ (lower dashed curve), and CPR of the fully ballistic constriction with $I(\phi)=(\pi G_T\Delta/e)\sin(\phi/2)\mathrm{sign}(\cos(\phi/2))$ (uppermost dashed curve), which is known as Kulik-Omelyanchuk formula [\onlinecite{KO}]. }
\label{fig-I-SNS}
\end{figure}

In the case of a large grain, $E_{\Th}\ll\Delta$, superconducting proximity effect is known to induce a mini-gap in the spectrum of the normal region~[\onlinecite{Golubov}]. Up to a numerical coefficient of the order of unity this gap is of the order of Thouless energy, $E_{g1}\simeq E_{\Th}$~[\onlinecite{Belzig}]. At energies just above that gap, $E-E_{g1}\ll E_{g1}$, the single particle density of states $\nu(E)$ has a universal square-root singularity $\nu(E)\propto \sqrt{E/E_{g1}-1}$~[\onlinecite{MV-1,MV-2}]. At finite superconducting phase  difference $\phi$ across the junction this gap feature closes at $\phi=\pi$ in accordance with the approximate formula $E_{g1}\simeq E_{\Th}|\cos(\phi/2)|$. At low temperatures, $T\ll E_{\Th}$, the resulting Josephson current as carried by Andreev sub-gap states is  almost perfectly harmonic~[\onlinecite{BB}]
\begin{equation}\label{I-SNS-1}
I(\phi)=\frac{G_TE_{\Th}}{e}\sin(\phi)\ln\left(\frac{2\Delta/E_{\Th}}{|\cos(\phi/2)|}\right)
\end{equation}
with weak logarithmic nonanalyticity stemming from the mini-gap feature. This formula remains valid even in the intermediate temperature regime, $E_{\Th}\ll T\ll\Delta$, with the only change that Thouless energy under the logarithm should be replaced by the temperature $T$. This anomalously weak temperature dependence should be contrasted to a conventional long SNS Josephson junctions, where raising the temperature above the excitation gap typically leads to an exponential suppression of the supercurrent $I(\phi)\simeq (G_TT/e)\sin(\phi)f(T/E_{\Th})$ with $f(z)=\sqrt{z}e^{-\sqrt{z}}$ for $z\gg1$~[\onlinecite{Zaikin,Dubos}]. 

In the opposite limit, $E_{\Th}\gg\Delta$, when superconducting proximity effect on the grain is strong and the induced spectral gap in the density of states reaches the value of $\Delta$, Josephson current was studied for several different models of NS interfaces. (\textit{i}) In the case of dirty tunneling barriers the zero-temperature Josephson CPR is found to be~[\onlinecite{BB,KL}] 
\begin{equation}\label{I-SINIS}
I(\phi)=\frac{G_T\Delta}{e}\sin(\phi)\mathbb{K}(\sin(\phi/2))
\end{equation}
where $\mathbb{K}(x)$ is the full elliptic integral of the first kind. The maximal critical current $I_c\simeq1.92G_T\Delta/e$ is achieved at the phase difference of $\phi_c\simeq1.18(\pi/2)$.  (\textit{ii}) In the case of disordered point contacts the supercurrent should be averaged over the distribution of transmission eigenvalues of the junction. This yields the Josephson current in the form~[\onlinecite{Beenakker-3UJE,KO,Heikkila}] 
\begin{equation}\label{I-SNS-2}
I(\phi)=\frac{\pi G_T\Delta}{e}\cos(\phi/2)\arcth(\sin(\phi/2))
\end{equation}
with only slightly higher critical current $I_c\simeq2.07G_T\Delta/e$. (\textit{iii}) In a chaotic cavity Josephson junction with identical ballistic contacts the supercurrents differ from Eq.~\eqref{I-SNS-2} because the distribution of transmission eigenvalues is different. One finds corresponding CPR in the form~[\onlinecite{BB}]
\begin{equation}\label{I-SCCS}
I(\phi)=\frac{4G_T\Delta}{e}\cot(\phi/2)[\mathbb{K}(\sin(\phi/2))-\mathbb{E}(\sin(\phi/2))]
\end{equation}
where $\mathbb{E}(x)$ is the complete elliptic integral of the second kind. All these types of Josephson junctions support parametrically the same critical current and their corresponding CPRs are plotted on Fig. \ref{fig-I-SNS} for the illustration. It is worth mentioning that Eqs.~\eqref{I-SNS-1}--\eqref{I-SNS-2} were originally derived based on the semicalssical theory of superconductivity from the Usadel equations; see review~[\onlinecite{GKI}] for the detailed discussion and references therein.     

\subsection{Motivation} 

It should be noted that Josephson currents as given by Eqs. \eqref{I-SINIS}--\eqref{I-SCCS} were essentially calculated in the quantum point contact limit of the junction, namely $\Delta/E_{\Th}\to0$. Naively one would expect that retaining $\Delta/E_{\Th}$ as a small, yet finite, parameter would not change these results considerably and give only subleading corrections to the current.  This is indeed the case for the magnitude of the critical current, which acquires a correction $\delta I_c/I_c\simeq -(\Delta/E_{\Th})\ln(E_{\Th}/\Delta)$~[\onlinecite{AL-Ic}]. There exists, however, a much more subtle effect that so far has received only very limited attention. Indeed, at finite $E_{\Th}$ the density of states in the metallic grain exhibits a nontrivial nonmonotonic behavior~[\onlinecite{AL-DOS}]. Remarkably, there exists a secondary gap, $E_{g2}$, that opens near the upper edge of the sub-gap spectrum close to $\Delta$ provided that Thouless energy is bigger than a certain threshold~[\onlinecite{Reutlinger-1,Reutlinger-2,Yokoyama}]. This double-gap feature in the proximity-induced DOS leads to a redistribution of the spectral current as carried by sub-gap Andreev states and ultimately renders the change in the shape of the Josephson CPR. We find a substantial skewed bending of the current at phases $\phi<\pi/2$ and a steeper fall-off of the current near the gap closing $\phi\to\pi$. 

Josephson junctions are being used as inductive elements for qubits. This motivated us to perform a detailed understanding of their CPRs, which is crucial in modeling of qubit nonlinearity. In light of the recent nanofabrication advances and development of gatemon qubits~[\onlinecite{Gatemon-1,Gatemon-2,Gatemon-3}] we focus our study on quantum dot Josephson junctions with multi-mode transparent interfaces. Coincidently, in this regime the effect of the secondary gap is the most pronounced. We also briefly discuss possible implications of our results for $\pi$-periodic Josephson circuits that support coherent transport of pairs of Cooper pairs (the ``$4e$" transport) and enable realization of protected qubits in rhombi chains [\onlinecite{Rhombi-1,Rhombi-2}].

\begin{figure}[b!]
\includegraphics[width=\linewidth]{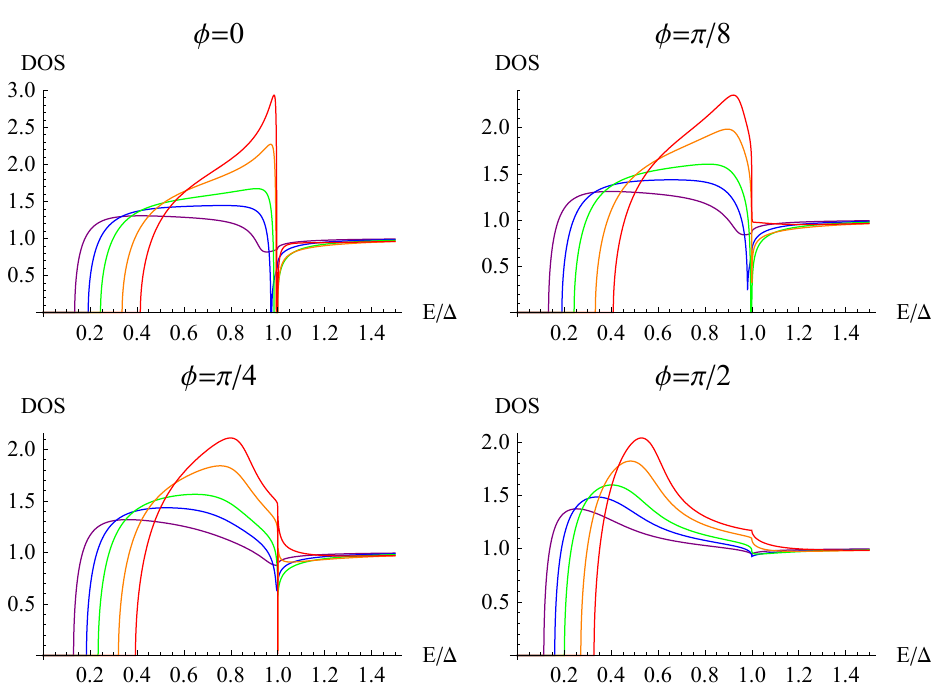} 
\caption{(Color online) Normalized proximity-induced density of states $\nu(E,\phi)/\nu_0$ in the quantum dot showing the usual gap $E_{g1}$ centered at $E=0$ and an additional secondary mini-gap $E_{g2}$ just below the gap edge $E=\Delta$. Different panels correspond to different phase across the junction $\phi=(0,\pi/8,\pi/4,\pi/2)$ while different lines on each panel correspond to different ratios between Thouless energy and superconducting gap: from the top line to the bottom one $E_{\Th}=(2\Delta, 1.5\Delta, \Delta, 0.75\Delta, 0.5\Delta)$.}
\label{fig-dos}
\end{figure}

\section{Formalism} 

We consider the normal diffusive grain/chaotic quantum dot with mean level spacing $\delta$ connected to two superconducting leads by quantum point contacts. We assume that the left(right) contacts are symmetric and have a large number of channels, $N_{1(2)}\gg 1$. We use the zero-dimensional version of the nonlinear $\sigma$-model to describe this system~[\onlinecite{Efetov,AL-KNLSM}]. The corresponding Keldysh action reads
\begin{align}\label{S}
S=-\frac{1}{2}\sum_{k=1,2}N_k\Tr\ln\left(1+\frac{T_k}{4}(\{\hat{G}_k,\hat{G}\}-2)\right)\nonumber\\ +i\delta^{-1}\Tr(E\hat{\tau}_3\hat{G})
\end{align}
The QD is described by the Green's function $\hat{G}$ which is a $4\times4$ matrix in the combined Keldysh and Nambu representation. We use two sets of Pauli matrices $\hat{\sigma}$ and $\hat{\tau}$ to distinguish these spaces respectively. The symbol of trace $\Tr(\ldots)$ implies all matrix summations and energy integration, while curly brackets $\{\hat{G}_k,\hat{G}\}$ under the trace denote matrix anti-commutator. The action is nonlinear because of the constraint $\hat{G}^2=1$. The first two terms in the sum of Eq. \eqref{S} represent coupling of the dot to the leads. The two superconducting reservoirs are assumed to have the same energy gap $\Delta$ and symmetric phase bias $\pm\phi/2$, so that the corresponding retarded/advanced Green's functions read 
\begin{equation}
\hat{G}^{R/A}_{1,2}= c^{R/A}_E\hat{\tau}_3+is^{R/A}_E[\hat{\tau}_1\cos(\phi/2)\pm\hat{\tau}_2\sin(\phi/2)],
\end{equation}
 where 
 \begin{equation}
c^{R/A}_E=\frac{-iE}{\sqrt{\Delta^2-(E\pm i0)^2}},\,\, s^{R/A}_E=\frac{\Delta}{\sqrt{\Delta^2-(E\pm i0)^2}} 
\end{equation}
for $E<\Delta$. To find the Green function $\hat{G}$ inside the grain one should solve the matrix saddle point equation for the action \eqref{S} which is given by the following commutator $[\hat{Q},\hat{G}]=0$, where $\hat{Q}=\hat{J}_1+\hat{J}_2-i\delta^{-1}E\hat{\tau}_3$ and 
\begin{equation}
\hat{J}_k=\frac{N_kT_k}{4+T_k(\{\hat{G}_k,\hat{G}\}-2)}\hat{G}_k.
\end{equation}
In this language, the single particle density of states is given by $\nu(E,\phi)=(\nu_0/2)\mathrm{Re}[\mathrm{tr}(\hat{\tau}_3\hat{G})]$, where $\nu_0$ is the density of states in the normal state and $\mathrm{tr}(\ldots)$ is the matrix trace without energy integration, whereas the current is given by $I(\phi)=(e/2\hbar)\Tr(\hat{\tau}_3\hat{\sigma}_3[\hat{J}_k,\hat{G}])$. The theory defined by the acton in Eq. \eqref{S} is equivalent to a more standard Keldysh-Green function formalism of Usadel equations commonly used in applications to SNS interferometers and double-barrier Josephson junctions [\onlinecite{Bezuglyi-1,Bezuglyi-2}].   

This formulation of the theory enables one to reproduce all the known special cases of Josephson junctions that we mentioned above. Indeed, the model of weakly-transparent tunneling contacts follows from Eq.~\eqref{S} by expanding the action at small transmissions $T_k\ll1$ and retaining only the linear term. Furthermore, neglecting the proximity effect on the normal region, thus replacing $\{\hat{G}_k,\hat{G}\}\to\{\hat{G}_1,\hat{G}_2\}$, one recovers the limit of superconductor-insulator-superconductor (SIS) junction with CPR $I(\phi)=(\pi G_T\Delta/2e)\sin(\phi)\tanh(\Delta/2T)$ 
as originally derived by Ambegaokar and Baratoff within the tunneling Hamiltonian approach [\onlinecite{AB}]. At temperatures close to the critical when superconducting energy gap is small, $\Delta\ll T$, this result further reduces to Aslamazov-Larkin formula [\onlinecite{AL}]: $I(\phi)=(\pi G_T\Delta^2/4eT)\sin(\phi)$ that was obtained earlier from the Ginzburg-Landau phenomenology.  Accounting for the proximity effect on the normal region, but still working in the limit of poorly transparent interfaces, namely retaining only the linear in $T_k$ term of the action \eqref{S}, $\propto T_k\{\hat{G}_k,\hat{G}\}$, one recovers CPR in the form of Eq. \eqref{I-SNS-1} for small Thouless energy.  In order to derive Eqs. \eqref{I-SINIS}-\eqref{I-SCCS} from Eq. \eqref{S} one has to keep arbitrary transmissions $T_k\in[0,1]$, and average the current $\int^{1}_{0}I(\phi)\rho(T_k)dT_k$ over the continuous distribution density of transmission eigenvalues $\rho(T_k)$. The latter takes a generic form [\onlinecite{Dorokhov,Schep-Bauer}]: $\rho(T_k)\propto (T^p_k\sqrt{1-T_k})^{-1}$ with normalization via the Landauer-Buttiker conductance $G_T=G_Q\int^{1}_{0}T_k\rho(T_k)dT_k$. The power exponent $p$ takes different values depending on the type of the contacts. For $p=3/2$, which corresponds to symmetric dirty interfaces with a high density of randomly distributed scatterers, namely SINIS type junction, one recovers Eq. \eqref{I-SINIS}. The case with $p=1$ corresponds to the Dorokhov function valid for a diffusive SNS connector and leads to CPR in the form of Eq. \eqref{I-SNS-2}. Lastly, the scenario with the power exponent $p =1/2$ corresponds to two ballistic connectors with equal conductances in series that translates to the current in the form of Eq. \eqref{I-SCCS}. Having in mind recently developed epitaxial Josephson junction devices, in this work we concentrate on the limit of fully transmitting channels, $T_k=1$, and allow for the arbitrary relationship between the Thouless energy and superconducting energy gap. 

\begin{figure}
\includegraphics[width=\linewidth]{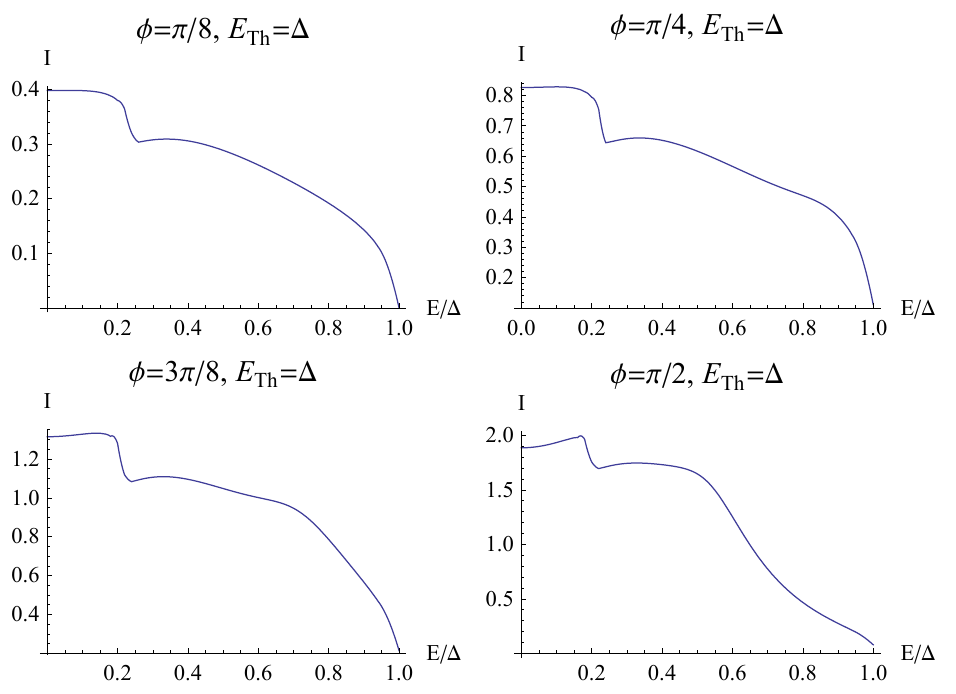} 
\caption{Energy-resolved spectral supercurrent of Andreev sub-gap states plotted in units of $I_0$ for several different values of the superconducting phase across the junction at $E_{\Th}=\Delta$.}
\label{fig-I-E}
\end{figure}

\section{Results} 

For completeness, we begin our discussion of main results with a brief recap of the behavior in the proximity-induced density of states. This analysis was exhaustively carried out in recent studies~[\onlinecite{Reutlinger-1,Reutlinger-2}] and served as a prerequisite for us to address the supercurrent. For sufficiently large Thouless energy, DOS displays rich sub-gap structure with the central gap $E_{g1}$ and the second gap $E_{g2}$ near $\Delta$, see Fig.~\ref{fig-dos}. At zero phase across the junction this second gap is estimated to be $E_{g2}\sim \Delta^3/E^2_{\Th}$ in the limit $E_{\Th}\gg\Delta$. For $E_{\Th}\sim\Delta$ the secondary gap is parametrically of the order $E_{\Th}$, yet it remains smaller than $E_{g1}$ in the same limit due to a numerical pre-factor. Gap $E_{g2}$ disappears below $E_{\Th}= 0.682\Delta$. Near the each gap edge DOS has a square-root singularity $\nu(E)\propto (E_{\Th}/\Delta)^2\sqrt{|E-E_g|/E_g}$, while at its maximum the DOS is of the order $\nu_{\mathrm{max}}\sim\nu_0(E_{\Th}/\Delta)$. At finite phase bias across the junction, the second gap closes at the critical phase $\phi_{c2}\simeq (\Delta/E_{\Th})$ whereas central gap closes at phase $\phi_{c1}=\pi$. The full phase dependence of $E_{g2}(\phi)$ was studied numerically, and was found to resemble the shape of a smile~[\onlinecite{Reutlinger-1}].

\begin{figure}[b!]
\includegraphics[width=\linewidth]{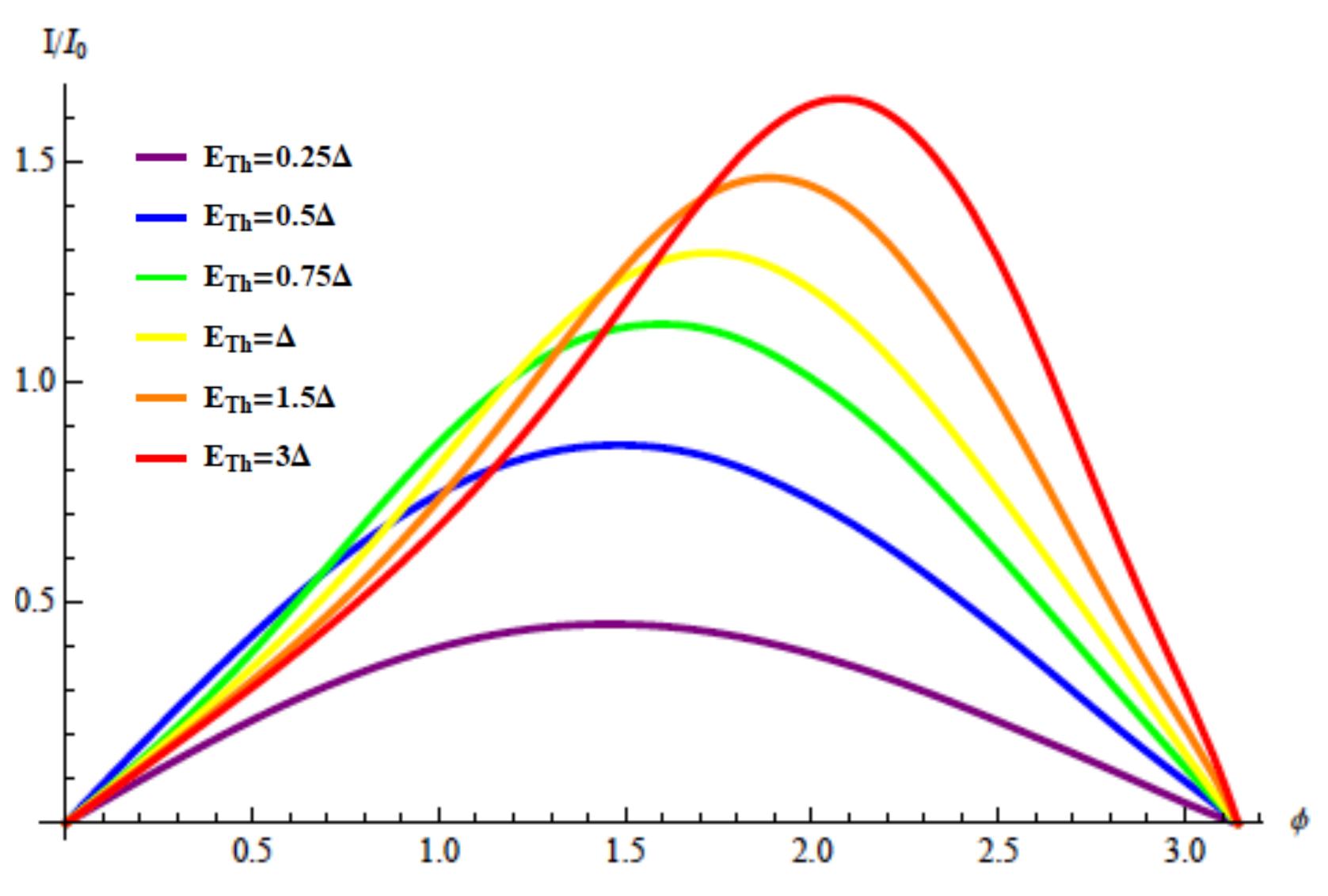} 
\caption{(Color online)  Josephson current-phase relationship $I(\phi)/I_0$ for a chaotic S-QD-S device with transparent interfaces: from the top line to the bottom one $E_{\Th}=(3\Delta, 1.5\Delta, \Delta, 0.75\Delta, 0.5\Delta, 0.25\Delta)$.}
\label{fig-I-CPR}
\end{figure}

\begin{figure*}
\includegraphics[width=0.95\linewidth]{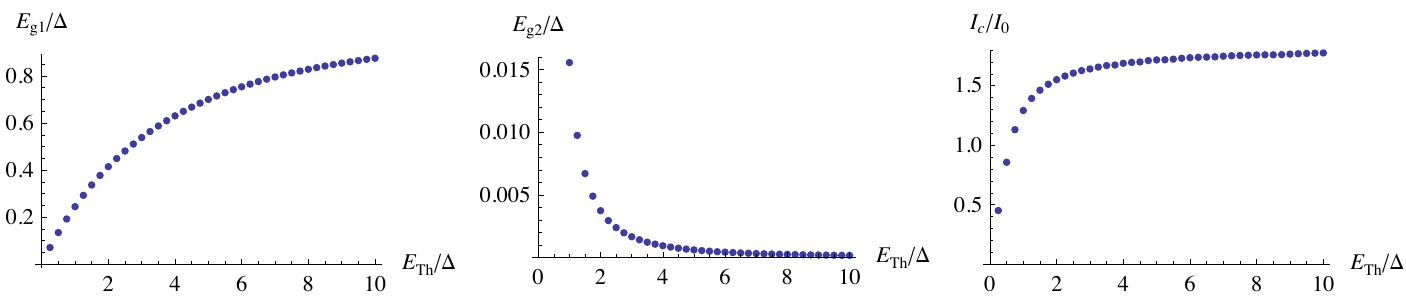} 
\caption{Crossover functions for the normalized energy gap, $E_{g1}/\Delta$ (on the left panel), the secondary gap $E_{g2}/\Delta$ (on the central panel), and the maximal critical current, $I_c/I_0$ (on the right panel), as a function of the ratio $E_{\Th}/\Delta$. The initial rise of $E_{g1}$ is linear in $E_{\Th}$, and the second gap can be fairly accurately approximated by $E_{g2}/\Delta\approx(17/2-6\sqrt{2})(\Delta/E_{\Th})^2$ for all values $E_{\Th}/\Delta>1$. The critical current also starts almost linearly with the Thouless energy as expected from Eq. \eqref{I-SNS-1}.}
\label{fig-Eg-Ic}
\end{figure*}

\begin{figure}
\includegraphics[width=\linewidth]{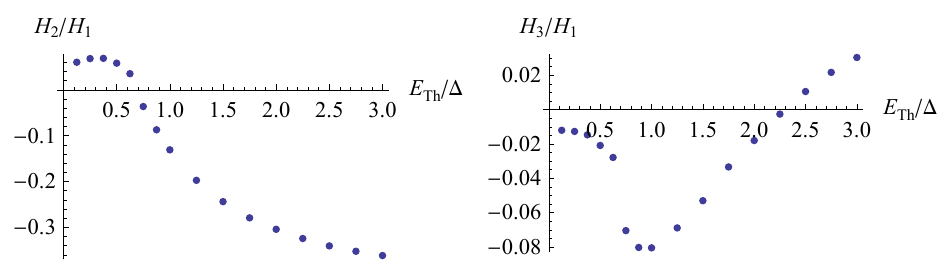} 
\caption{Ratios of the second harmonic, $H_2$, on the left panel, and the third harmonic, $H_3$, on the right panel, of the Josephson current to its first harmonic, $H_1$, presented for different values of the Thouless energy versus energy gap.}
\label{fig-Hn}
\end{figure}

This complicated sub-gap behavior changes the spectral flow of the supercurrent. To see this clearly, it is useful to plot the energy-resolved current of Andreev states. This is shown in Fig. \ref{fig-I-E} for zero temperature with several different phases, and a choice of $E_{\Th}=\Delta$. One should notice a pronounced kink in the function that correlates with the edge of a minigap. A second smaller kink develops close to the energy $\Delta$ for higher values of $E_{\Th}$ which is a manifestation of the second gap. By integrating the spectral current over all states one finds the Josephson CPR. We highlight the resulting curves in Fig. \ref{fig-I-CPR}. The effect of bending in the current at phases $\phi<\pi/2$ that starts to develop at $E_{\Th}>\Delta$ we primarily attribute to a formation of sub-gaps in the density of states.  To make a clear connection between the energy gap and critical current we studied how they saturate as a function of Thouless energy. These results are presented on Fig.~\ref{fig-Eg-Ic}. To characterize the CPR curves further we introduce Fourier components, $H_n=(2/\pi)\int^{\pi}_{0}I(\phi)\sin(n\phi)d\phi$, and plot them for different ratios of $E_{\Th}/\Delta$, see Fig.~\ref{fig-Hn}. We notice that the second harmonic changes sign near $E_{\Th} = 0.7\Delta$. Perhaps more interestingly, we observe that the magnitude of the second harmonic becomes comparable to the first harmonic at large Thouless energy. The third harmonic changes sign near $E_{\Th} = 2.3\Delta$ while its magnitude remains small compared to the second harmonic. It is also of interest to look at the Taylor coefficients of the current $C_{n}=1/(n!)\lim_{\phi\to0}\partial^n_\phi (I(\phi)/I_c)$. In particular, $C_3$ controls the nonlinearity of the Josephson device qubit that can be modified by applying gate voltages to a junction that would change $E_{\Th}$. The initial almost linear slope of $C_3$ versus $E_{\Th}/\Delta\ll1$, can be readily seen from Eq.~\eqref{I-SNS-1}, with asymptotic behavior $C_3\propto E_{\Th}/\Delta$. In the opposite limit, $C_3$ saturates to a constant that is numerically close to $C_3\simeq 0.05$. The general form of $C_3(E_{\Th}/\Delta)$ is a complicated nonmonotonic function.     

\section{Discussions}  

A few comments are in order in relation to results presented in this paper. The hard gap features in the density of states correspond only to the mean field level (saddle point) treatment of the action Eq.~\eqref{S}. Fluctuations (instantons) on top of the saddle point will give raise to the Lifshitz-type tail-states below the gap~[\onlinecite{Meyer,Ostrovsky}]. Mathematically it bears a close analogy with the Tracy-Widom distribution for the DOS tail in the random matrix theory (RMT)~[\onlinecite{MV-1}]. Indeed, in the regime of the mini-gap, $E_{\Th}\ll\Delta$, the asymptotic behavior of DOS close to the central gap edge is $\ln\nu(E)\simeq -g(1-E/E_{g1})^{3/2}$ for $E_{g1}-E\ll E_{g1}$, where $g\gg1$ is the dimensionless conductance of the N region. In the deep low-energy limit, $E\ll E_{g1}$, the behavior of the DOS is log-normal $\ln\nu(E)\simeq-g\ln^2(E_{g1}/E)$. We expect similar tail-states to exist in the region of the second gap $E\sim E_{g2}$ although the functional form of their scaling close to the gap edge may be different. 

The calculation of the supercurrent was carried out here for the ensemble-averaged Green's function, and can not therefore describe the mesoscopic fluctuations of $I(\phi)$ from the average. These fluctuations are known to be universal in the regime $E_{\Th}\gg\Delta$ where variance of the current scales as $\mathrm{var}\{ I(\phi)\}\propto(e\Delta/h)^2$~[\onlinecite{Beenakker-3UJE,Chalker}]. This can be immediately concluded from the circuit theory knowing that the supercurrent $I(\phi)$ is a linear statistic on transissions $T_k$. In the opposite limit, $E_{\Th}\ll\Delta$, fluctuations are not universal and scale with Thouless energy $\mathrm{var}\{I(\phi)\}\propto (eE_{\Th}/h)^2$~[\onlinecite{Spivak,Micklitz,Houzet}].  Another important mesoscopic coherence effect is that of weak localization. Such corrections to supercurrent are known to be small in inverse dimensionless conductance of the normal region, $\delta I_c/I_c\sim 1/g\ll1$,  irrespective of the relationship between the Thouless energy and the superconducting gap~[\onlinecite{Houzet}].  
            
Finally, we discuss two aspects of the CPR of Josephson junctions in light of their use in qubits. It has been recently proposed~[\onlinecite{Rhombi-1,Rhombi-2}] that special Josephson elements whose first harmonic of Josephson energy $V(\phi)=E_{J1}\cos(\phi)+E_{J2}\cos(2\phi)$ is suppressed compared to the second harmonic may realize protected qubits against charge noise as transfer of single Cooper pairs is strongly suppressed in such devices. An effective $\cos(2\phi)$ element can be formed by placing two such junctions in parallel and biasing the resulting loop with external flux to suppress the first harmonic of the Josephson energy. Two such plaquettes form a minimal protected element. If $N\gg 1$ is the total number of plaquettes in the qubit, the suppression of sensitivity to local noise due to inevitable static disorder in the parameter values of the device is then exponential $\exp(-N \ln(E_{J2}/E_{J1}))$. Our calculations of supercurrent in S-QD-S circuits reveal that even such basic junctions can realize desirable properties of current-phase relationship with comparable magnitudes of amplitudes in current harmonics. Lastly, we wish to comment on the anharmonicity of gatemon-qubit as recently realized in epitaxial InAs-Al junctions~[\onlinecite{Gatemon-3}]. For a multimode junction qubit Josephson energy was modeled by the usual circuit theory expression $V(\phi)=-\Delta\sum_k\sqrt{1-T_k\sin^2(\phi/2)}$. By expanding energy over phase and retaining the first two terms one gets a harmonic oscillator and its quartic nonlinearity. The latter translates into the anharmonicity of the qubit as quantified by a parameter $\alpha\approx -E_C[1-(3/4)\sum_kT^2_k/\sum_kT_k]$, where $E_C$ is the charging energy. For fully transmitting channels $\alpha=-E_C/4$. As we have shown, Josephson CPR of the multimode junction with transparent interfaces deviates from the prediction of the circuit theory because it does not account properly for the intricate details of the superconducting proximity effect on the normal region. Thus more accurate theoretical modeling of gatemon-qubit nonlinearities remains an open task, and our theory will be useful for that purpose.                 

\subsection*{Acknowledgment} 

We thank Dushko Kuzmanovski for developing an initial code of the computer program that served the basis for numerical results presented in this study. We thank Norman Birge for reading the paper and providing suggestions for improvements. This work was financially supported in part by the NSF CAREER Grant No. DMR-1653661, NSF EAGER Grant No. DMR-1743986, and by the DOE Office of Science, under Award Number DE-SC0017888 (A.L. and C.W.); by the Army Research Office Grant W911NF-15-1-0248, and the Wisconsin Alumni Research Foundation (M.V. and A.L.). 
  

\end{document}